\documentclass{PoS}
\usepackage{hepnames}
\usepackage{ragged2e}
\usepackage{hepunits}

\title{  Proper time evolution of magnetic susceptibility in a magnetized plasma of quarks and gluons  }

\ShortTitle{Proper time evolution of magnetic susceptibility in a magnetized quark-gluon plasma}

\author{\speaker{S.M.A. Tabatabaee} and N. Sadooghi \\
        Department of Physics, Sharif University of Technology, P.O. Box 11155-9161, Tehran, Iran\\
 E-mail: \email{tabatabaeemehr\_sma@physics.sharif.ir, sadooghi@physics.sharif.ir}
}

\abstract{In ultrarelativistic heavy ion collisions, enormous magnetic fields are generated because of fast-moving charged particles. In the presence of these magnetic fields, the spin of particles is aligned either in the parallel or in the antiparallel direction with respect to the direction of the magnetic field. A finite magnetization is thus produced. It is known that a finite magnetic susceptibility, $\chi_{m}$, changes the evolution of the energy density of the quark-gluon plasma (QGP), which is believed to be created in these collisions. Depending on whether the system under consideration is a paramagnetic ($\chi_{m}>0$) or diamagnetic ($\chi_{m}<0$) fluid, it slows down or speeds up the decay of the energy density, and affect other thermodynamic quantities.
In general, one expects that the magnetic susceptibility depends on the magnetic field and temperature. Bearing in mind that these parameters evolve with the evolution of the fluid, a nonuniform magnetic susceptibility in this system is thus expected. In this work, we first determine $\chi_{m}$ by using a certain analogy to the standard anisotropic kinetic theory, where the one-particle distribution function is replaced by the corresponding anisotropic distribution function. We then determine the proper time dependence of the magnetic susceptibility in the framework of the ideal transverse magnetohydrodynamics. We also study the effect of dissipation on the evolution of $\chi_{m}$.}

\FullConference{XIII Quark Confinement and the Hadron Spectrum - Confinement2018\\
		31 July - 6 August 2018\\
		Maynooth University, Ireland}

\begin{document}
%%%%%%%%%%%%%%%%%%%%%%%%%%%%%%%%
\section{Introduction}
%%%%%%%%%%%%%%%%%%%%%%%%%%%%%%%%
The recently observed polarization of $\Lambda$-particles at the Relativistic Heavy Ion Collider (RHIC) is related to the coupling of the spin of the constituent particles to the angular momentum of the fluid \cite{STAR:2017ckg}. This is, however, not the only source for the spin polarization. In these kinds of collisions, extremely large magnetic fields are believed to be generated \cite{Kharzeev:2007jp}. It is thus possible that the interaction of these fields with the spin of particles has an imprint on the polarization of particles. To study the effect of magnetic field on this polarization, the anisotropy caused by a finite magnetization and its evolution during the hydrodynamical stage are to be taken into account. In what follows, after briefly introducing the magnetohydrodynamic (MHD) setup, we use an analogy to anisotropic hydrodynamics, and determine, in particular, the proper time evolution of the magnetic susceptibility by making use of a modified anisotropic one-particle distribution function in the framework of the kinetic theory.
%%%%%%%%%%%%%%%%%%%%
\section{MHD setup}
%%%%%%%%%%%%%%%%%%%%
In the presence of external magnetic fields, an ideal magnetized fluid is described by a total energy-momentum tensor \cite{Israel:1978up}
\begin{eqnarray} \label{2.1}
T^{\mu \nu} &=& T^{\mu \nu}_{f} + T^{\mu \nu}_{em},
\end{eqnarray}
where the fluid and electromagnetic energy-momentum tensors, $T^{\mu\nu}_{f}$ and $T^{\mu\nu}_{em}$, are given by
\begin{eqnarray}\label{2.2}
 T^{\mu \nu}_{f} &=& (\epsilon + p ) u^{\mu}u^{ \nu} - p g^{\mu \nu} - \Pi^{(\mu}u^{\nu)} + F^{(\mu}_{~~~\alpha}M^{\nu) \alpha} \nonumber \\
T^{\mu \nu}_{em} &=& F^{\mu}_{~~~\lambda}F^{\lambda \nu} + \frac{1}{4} g^{\mu \nu} F_{\alpha \beta} F^{\alpha \beta}.
\end{eqnarray}
Here, $u^{\mu}, \epsilon$ and $p$ are the fluid four velocity, energy density, and  pressure, $F_{\mu\nu}$ and $M_{\mu\nu}$ are the field strength and magnetization tensors. Moreover, the auxiliary vector $\Pi^{\mu}$ is defined by $\Pi^{\mu}\equiv 2 u_{\lambda} F^{[\mu}_{~~~\nu}M^{\lambda] \nu}$. For the sake of brevity, we use the symmetrization and antisymmetrization symbols, $A^{(\mu \nu)} \equiv \frac{1}{2} \left( A^{\mu \nu}+A^{\nu \mu}  \right)$ and $A^{[\mu \nu]} \equiv \frac{1}{2} \left( A^{\mu \nu} - A^{\nu \mu}  \right)$.
Assuming an ideal fluid with infinite conductivity, we neglect, in what follows, the electric field. In this case, the field strength tensor and magnetization are expressed in terms of the magnetic field as
\begin{eqnarray} \label{2.3}
F^{\mu \nu} &=& - B b^{\mu \nu} , \quad M^{\mu \nu} = -  M b^{\mu \nu},
\end{eqnarray}
with $b^{\mu \nu} \equiv \epsilon^{ \mu \nu \alpha \beta } b_{\alpha} u_{\beta}$ and $b^{\mu} \equiv \frac{B^{\mu}}{B}$. The magnetic four-vector is thus given by  $B^{\mu} \equiv \frac{1}{2} \epsilon^{\mu \nu \alpha \beta} F_{\nu \alpha} u_{\beta}$, and is normalized as $B^{\mu} B_{\mu} = - B^{2}$.
Once the electric field is assumed to be infinitely small, $\Pi^{\mu}$ from (\ref{2.2}) vanishes. In the absence of the auxiliary vector, $T_{f}^{\mu \nu}$ is expressed as
\begin{eqnarray}\label{2.4}
T^{\mu \nu}_{f} &=& (\epsilon + p_{\perp} ) u^{\mu} u^{\nu} - p_{\perp} g^{\mu \nu} + (p_{\|}-p_{\perp}) b^{\mu}b^{ \nu},
\end{eqnarray}
with $p_{\perp} = p-BM$ and $p_{\|}=p$. In this forms, $T_{f}^{\mu\nu}$ from \eqref{2.4} is comparable with the energy-momentum tensor
\begin{eqnarray} \label{2.5}
T_{f}^{\mu\nu}=(\epsilon+p)u^{\mu}u^{\nu}-p_{T}g^{\mu\nu}+\left(p_{L}-p_{T}\right)z^{\mu}z^{\nu},
\end{eqnarray}
that is introduced in the context of anisotropic hydrodynamics \cite{strickland2010}. The only difference is that in (\ref{2.5}), the subscripts $T$ and $L$ correspond to transverse and longitudinal direction with respect to the anisotropy direction $\hat{e}_{z}$, while in (\ref{2.4}), the subscripts $\perp$ and $\|$ correspond to transverse and longitudinal directions with respect to the direction of the magnetic field. Moreover, whereas the anisotropy in (\ref{2.5}) arises due to viscous effects, in (\ref{2.4}) anisotropies originate from the magnetization of fluid.
\par
In the ideal MHD, the dynamics of the fluid and electromagnetic field is solely described by the energy-momentum conservation, $\partial_{\mu} T^{\mu \nu}=0$, supplemented with the Maxwell equations,
\begin{eqnarray}\label{2..6}
\partial_{\mu}\tilde{F}^{\mu\nu}=0,\quad\mbox{and}\quad \partial_{\mu}F^{\mu\nu}=J^{\nu}.
\end{eqnarray}
Here, the dual field strength tensor is given by
\begin{eqnarray}\label{2.7}
\tilde{F}^{\mu\nu}=B^{\mu}u^{\nu}-B^{\nu}u^{\mu}.
\end{eqnarray}
Assuming the magnetic field to be initially aligned in the $z$-direction, and neglecting the transverse expansion of the fluid, the homogeneous Maxwell equation leads to the proper time evolution of the magnitude of the magnetic field \cite{Pu:2016ayh, shokri-1}
\begin{eqnarray}\label{2.8}
B=B_{0} \left( \frac{\tau_{0}}{\tau} \right),
\end{eqnarray}
with the proper time $\tau\equiv (t^2-y^2)^{1/2}$. Here, $B_{0}=B(\tau_{0})$ is the magnetic field at the initial time $\tau_0$.
To derive \eqref{2.8}, the Bjorken flow $u^{\mu}=\gamma (1,0,v_y,0)$ with  $v_{y}=y/t$ is used. As it is shown in \cite{Pu:2016ayh, shokri-1}, \eqref{2.8} is a manifestation of the frozen flux theorem in the assumed ideal transverse MHD setup, which also guarantees that the direction of the magnetic field is frozen, and does not evolve with the expansion of the fluid.
%%%%%%%%%%%%%%%%%%%%%%%%%%%%%%%%
\section{Anisotropic MHD description of a nondissipative fluid}
%%%%%%%%%%%%%%%%%%%%%%%%%%%%%%%%
To construct an energy-momentum tensor $T^{\mu\nu}_{0}=\int d\tilde{k}~ k^{\mu}k^{\nu}f_{0}$ which incorporates the magnetization, we introduce the one-particle distribution function \cite{A}
\begin{eqnarray}\label{3.1}
f_{0}=\exp\left(-\sqrt{k^{\mu}\Xi_{\mu\nu}^{(0)}k^{\nu}}/\lambda_{0}\right).
\end{eqnarray}
Here, $d\tilde{k}\equiv \frac{d^{3}k}{(2\pi)^{3}|\boldsymbol{k}|}$ and $k^{\mu}$ and $\lambda_{0}$ are the four-momentum of particles and the effective temperature, respectively. Moreover, the anisotropic tensor $\Xi_{\mu\nu}^{(0)}$ is given by
\begin{eqnarray}\label{3.2}
\Xi_{\mu\nu}^{(0)}\equiv u_{\mu}u_{\nu}+\xi_{0} b_{\mu}b_{\nu},
\end{eqnarray}
where $\xi_{0}$ is the dynamical anisotropy parameter, induced by the magnetization of the fluid. In the ideal transverse MHD, $\lambda_{0}$ and $\xi_{0}$ depend, in general, on the proper time $\tau$ and rapidity $\eta$ defined by $\eta\equiv \frac{1}{2}\ln\frac{t+y}{t-y}$ in transverse MHD. However, by the assumption of boost invariance, they become $\eta$ independent. In the presence of an external magnetic field, $f_{0}$ satisfies the Boltzmann equation
\begin{eqnarray}\label{3.3}
k^{\mu}\partial_{\mu}f_{0}+q_{f}eF^{\mu\nu}k_{\nu}\frac{\partial f_{0}}{\partial k^{\mu}}=C[f_{0}],
\end{eqnarray}
with $q_{f} e $ being the charge of various flavors $f$. Using the relaxation time approximation (RTA), we set, as in \cite{strickland2017},
\begin{eqnarray}\label{3.4}
C[f_{0}]=-(k\cdot u)\left(\frac{f_{0}-f_{eq}}{\tau_{r,0}}\right),
\end{eqnarray}
with the equilibrium one-particle distribution function given by
\begin{eqnarray}\label{3.5}
f_{eq}=\exp\left(-(k\cdot u)/T\right),
\end{eqnarray}
and the relaxation time $\tau_{r,0}$. In \cite{strickland2017}, the relaxation time is brought in connection with the shear viscosity over the entropy density ratio. In ideal MHD discussed here, however, the fluid is dissipationless. The relaxation time $\tau_{r,0}$ is thus only related to the magnetization of the fluid, that, because of the induced anisotropy, affects $f_{eq}$.
%%%%%%%%%%%%%%%%%%%%%%%%%%%%%%%%
\subsection*{Evolution of $\xi_0$ and $\lambda_0$}\label{sec3.1}
%%%%%%%%%%%%%%%%%%%%%%%%%%%%%%%%
To obtain the evolution equation for $\xi_0$ and $\lambda_0$, we use the method of moments of Boltzmann equation \eqref{3.3}. The zeroth and first moment of the Boltzmann equation lead to two coupled differential equation for $\xi_0$ and $\lambda_0$,
\begin{eqnarray}\label{3.7}
\frac{\partial_{\tau}\xi_{0}}{1+\xi_{0}}-\frac{6\partial_{\tau}\lambda_{0}}{\lambda_{0}}-\frac{2}{\tau}&=&\frac{2}{\tau_{r,0}}\left(1-{\cal{R}}^{3/4}(\xi_{0})\sqrt{1+\xi_{0}}\right),\nonumber\\
\frac{\partial {\cal{R}}(\xi_{0})}{\partial\xi_0}\frac{\partial_{\tau}\xi_0}{{\cal{R}}(\xi_{0})}+\frac{4\partial_{\tau}\lambda_0}{\lambda_0}&=&-\frac{1}{2\tau\xi_0}\left(3\xi_0-1+\frac{1}{(1+\xi_0){\cal{R}}(\xi_{0})}\right),
\end{eqnarray}
in analogy to the differential equations arising in anisotropic hydrodynamics in \cite{strickland2017}. In \eqref{3.7},
\begin{eqnarray}\label{3.8}
{\cal{R}}(\xi)\equiv\frac{1}{2}\left(\frac{1}{1+\xi}+\frac{\mbox{tan}^{-1}\sqrt{\xi}}{\sqrt{\xi}}\right).
\end{eqnarray}
The solution of these coupled differential equations leads to the proper time evolution of all thermodynamic quantities through their dependence on $\xi_0$ and $\lambda_0$ (for more details, see \cite{A}).
%%%%%%%%%%%%%%%%%%%%%%%%%%%%%%%%
\section{Anisotropic MHD description of a dissipative fluid}
%%%%%%%%%%%%%%%%%%%%%%%%%%%%%%%%
%%%%%%%%%%%%%%%%%%%%%%%%%%%%%%%%%%%
\subsection*{Determination of the dissipative part of the one-particle distribution function, shear, and bulk viscosities}\label{sec4a}
%%%%%%%%%%%%%%%%%%%%%%%%%%%%%%%%%%%
To determine the dissipative part of the one-particle distribution function, we start by plugging $f=f_b+\delta f_d$, including the nondissipative and dissipative one-particle partition functions, $f_b$ and $\delta f_{d}$, into the Boltzmann equation
\begin{eqnarray}\label{4.1}
k^{\mu}\partial_{\mu}f+q_{f}eF^{\mu\nu}k_{\nu}\frac{\partial f}{\partial k^{\mu}}=C[f].
\end{eqnarray}
Here, as in the previous section, $f_{b}$ is given by
\begin{eqnarray}\label{4.2}
f_{b}\equiv \exp\left(-\sqrt{k^{\mu}\Xi_{\mu\nu}k^{\nu}}/\lambda\right),
\end{eqnarray}
with $
\Xi_{\mu\nu}=u_{\mu}u_{\nu}+\xi b_{\mu}b_{\nu}$. Moreover, $\lambda$ and $\xi$ are the effective temperature and anisotropy parameters, that are, in contrast to $\lambda_{0}$ and $\xi_{0}$ from previous section, affected by the dissipation of the fluid. To determine $\delta f_{d}$, we follow the method that has been used in \cite{roy2018}. We use rank-two tensors
\begin{eqnarray}\label{4.7}
U_{\mu\nu}^{(0)}\equiv \Delta_{\mu\nu}, \quad U_{\mu\nu}^{(1)}\equiv b_{\mu}b_{\nu}, \quad U_{\mu\nu}^{(2)}\equiv b_{\mu\nu},
\end{eqnarray}
with $\Delta_{\mu\nu}\equiv g_{\mu\nu}-u_{\mu}u_{\nu}$, and traceless
\begin{eqnarray}\label{4.4}
\begin{array}{rclcrcl}
V^{(0)} &=& \xi^{(2)}-\frac{2}{3}\xi^{(1)},&\quad&
V^{(1)} &=&\xi^{(2)}-\xi^{(1)} - \xi^{(3)} + \xi^{(4)} + \xi^{(5)},  \\
V^{(2)} &=& - \left( \xi^{(5)} + 4 \xi^{(4)} \right), &\quad&V^{(3)} &=&\xi^{(6)}+\xi^{(7)}, \\
V^{(4)} &=& \xi^{(7)},&\quad&&&\\
\end{array}
\end{eqnarray}
as well as traceful rank-four tensors
\begin{eqnarray}\label{4.5}
W^{(0)}= \xi^{(1)}, \quad \mbox{and}\quad
W^{(1)}= \xi^{(3)},
\end{eqnarray}
with $\xi_{\mu\nu\rho\sigma}^{(i)}, i=1,\cdots 7$ given by \cite{kineticbook}
\begin{eqnarray}\label{4.6}
\begin{array}{rclcrcl}
\xi_{\mu\nu\rho\sigma}^{(1)} &=& \Delta_{\mu \nu} \Delta_{\rho \sigma},   &\qquad&
\xi_{\mu\nu\rho\sigma}^{(2)} &=&  \Delta_{\mu \rho} \Delta_{\nu \sigma} + \Delta_{\mu \sigma} \Delta_{\nu \rho},\\
\xi_{\mu\nu\rho\sigma}^{(3)} &=& \Delta_{\mu \nu} b_{\rho} b_{\sigma} + \Delta_{\rho \sigma} b_{\mu} b_{\nu},&\qquad&
\xi_{\mu\nu\rho\sigma}^{(4)} &=& b_{\mu} b_{\nu} b_{\rho} b_{\sigma} ,  \\
\xi_{\mu\nu\rho\sigma}^{(5)} &=& \Delta_{\mu \rho} b_{\nu} b_{\sigma} + \Delta_{\nu \rho} b_{\mu} b_{\sigma} + \Delta_{\mu \sigma} b_{\rho} b_{\nu} + \Delta_{\nu \sigma} b_{\rho} b_{\mu},\\
\xi_{\mu\nu\rho\sigma}^{(6)} &=& \Delta_{\mu \rho} b_{\nu \sigma}  + \Delta_{\nu \rho} b_{\mu \sigma}  + \Delta_{\mu \sigma}  b_{\nu \rho} + \Delta_{\nu \sigma} b_{\mu \rho},&&&& \\
\xi_{\mu\nu\rho\sigma}^{(7)} &=& b_{\mu \rho} b_{\nu} b_{\sigma} + b_{\nu \rho} b_{\mu} b_{\sigma} + b_{\mu \sigma} b_{\rho} b_{\nu} + b_{\nu \sigma} b_{\rho} b_{\mu},&&&&
\end{array}
\end{eqnarray}
and arrive after a lengthy but straightforward computation at the final expression for $\delta f_{d}$,
\begin{eqnarray}\label{4.8}
\delta f_{d}=\sum_{n=0}^{2} \ell_{n}k^{\mu}k^{\nu}U_{\mu\nu}^{(n)}+\sum_{n=0}^{4}g_{n}
k^{\mu}k^{\nu}V_{\mu\nu}^{(n)}+\sum_{n=0}^{1}h_{n}k^{\mu}k^{\nu}W_{\mu\nu}^{(n)}-(f_{b}-f_{eq}).
\end{eqnarray}
Here, $U_{\mu\nu}^{(n)}, n=0,1,2$ is given in \eqref{4.7}, and
\begin{eqnarray}\label{4.8a}
V_{\mu\nu}\equiv V_{\mu\nu\rho\sigma}^{(n)}\omega^{\rho\sigma},\qquad
W_{\mu\nu}\equiv W_{\mu\nu\rho\sigma}^{(n)}\omega^{\rho\sigma},
\end{eqnarray}
are definied in terms of rank-four tensors $V^{(n)}_{\mu\nu\rho\sigma}$ and $W^{(n)}_{\mu\nu\rho\sigma}$ from \eqref{4.4} and \eqref{4.5}, as well as $\omega^{\mu\nu}\equiv \frac{1}{2}\left(\nabla_{\mu}u_{\nu}+\nabla_{\nu}u_{\mu}\right)$ with $\nabla_{\mu}\equiv\Delta_{\mu\rho}\partial^{\rho}$.
The coefficients $\ell_{n}, g_n$, and $h_{n}$ in \eqref{4.8} are given by
\begin{eqnarray}\label{4.8a}
\ell_{0}=\nu_{H} \frac{D \lambda}{\lambda}, \quad \ell_{1} = \nu_{H}\xi\left(\frac{D \xi}{\xi} -\frac{D \lambda}{\lambda} \right), \quad \ell_{2} = 0,
\end{eqnarray}
where $D\equiv u_{\mu}\partial^{\mu}$, $\nu_{H} \equiv \frac{\tau_{r} f_{b}}{\lambda (k\cdot u) H_{b}}$ with $H_{b}(k) \equiv \sqrt{1 + \xi \frac{(k \cdot b)^{2}}{( k \cdot u)^{2}}}$,
\begin{eqnarray}\label{4.9}
g_{0}=  \frac{\nu_{H}}{2} ,\quad
g_{1}=-  \frac{2\nu_{H}\chi_{H}^{2}}{(1+4\chi_{H}^{2})},\quad
g_{2}=-  \frac{\nu_{H}\chi_{H}^{2}}{2(1+\chi_{H}^{2})},\quad
g_{3}= - \frac{\nu_{H}\chi_{H}}{2(1+4\chi_{H}^{2})},\quad
g_{4}=\frac{\nu_{H}\chi_{H}}{2(1+\chi_{H}^{2})},\nonumber\\
\end{eqnarray}
with $\chi_{H} \equiv \frac{q_{f} e B \tau_{r}}{k\cdot u}$, and
\begin{eqnarray}\label{4.10}
h_{0}=\frac{\nu_{H}}{3}, \qquad\mbox{and}\qquad h_{1} = 0.
\end{eqnarray}
Plugging, at this stage,
$\delta f_{d}$ from \eqref{4.8} into
$
\tau^{\mu\nu}=\int d\tilde{k}~k^{\mu}k^{\nu}\delta f_{d},
$
and comparing the resulting expression with
\begin{eqnarray}\label{4.12}
\tau_{\mu\nu}&=&\sum\limits_{n=0}^{1}\alpha_{n}U^{(n)}_{\mu\nu}+
\sum_{n=0}^{4}\eta_{n}V^{(n)}_{\mu\nu}+
\sum_{n=0}^{1}\tilde{\zeta}_{n}W^{(n)}_{\mu\nu} -(T^{b}_{\mu\nu}-T^{eq}_{\mu\nu}),
\end{eqnarray}
the dissipative part of energy-momentum tensor, $\tau_{\mu\nu}$ is determined. In \eqref{4.12}, $T_{i}^{\mu\nu}\equiv \int d\tilde{k}k^{\mu}k^{\nu}f_{i}, i=b,eq$, and
\begin{eqnarray}\label{4.13}
\alpha_{n}&=&\frac{1}{3}\int d\tilde{k}~\ell_{n}|\boldsymbol{k}|^{4}, \quad  \eta_{n}=\frac{2}{15}\int d\tilde{k}~g_{n}|\boldsymbol{k}|^{4}, \quad \tilde{\zeta}_{n}=\frac{1}{3}\int d\tilde{k}~h_{n}|\boldsymbol{k}|^{4},
\end{eqnarray}
are given in terms of $\ell_{n}$, $g_{n}$ and $h_{n}$ from \eqref{4.8a}-\eqref{4.10}. They are related to the shear and bulk viscosity of the fluid (see \cite{A} for more details).
%%%%%%%%%%%%%%%%%%%%%%%%%%%%%%%%
\subsection*{Evolution of $\xi$ and $\lambda$}
%%%%%%%%%%%%%%%%%%%%%%%%%%%%%%%%
To arrive at the corresponding differential equations for $\xi$ and $\lambda$, we use, as in Sec. \ref{sec3.1}, the zeroth and first moment of Boltzmann equation \eqref{4.1} in the RTA,
\begin{eqnarray}\label{4.x}
C[f]=-(k\cdot u)\left(\frac{f-f_{eq}}{\tau_{r}}\right),
\end{eqnarray}
where $f_{eq}$ is defined in \eqref{3.5} and $\tau_{r}$ is the relaxation time in a dissipative fluid. After some work, we arrive at two coupled differential equations
\begin{eqnarray}\label{4.15}
&&\hspace{-0.3cm}\frac{\partial_{\tau}\xi}{1+\xi}-\frac{6\partial_{\tau}\lambda}{\lambda}-\frac{2}{\tau}=\frac{2}{\tau_{r}}\left(1-{\cal{R}}^{3/4}(\xi)\sqrt{1+\xi}\right). \nonumber\\
&&\hspace{-0.3cm}\frac{4\partial_{\tau}\lambda}{\lambda}\bigg[1-\frac{\tau_{r}}{12\tau}\left(3+\frac{1}{{\cal{R}}(\xi)(1+\xi)^{2}}\right)\bigg]+\frac{\partial_{\tau}\xi}{{\cal{R}}(\xi)}\frac{\partial {\cal{R}}(\xi)}{\partial\xi}+\frac{4}{3\tau}-\frac{2\tau_{r}}{15}\left(3+\frac{1}{{\cal{R}}(\xi)(1+\xi)^{2}}\right)\frac{1}{\tau^2}=0,\nonumber\\
\end{eqnarray}
whose solution leads to the $\tau$ dependence of anisotropy parameter and effective temperature in the dissipative case. Here, ${\cal{R}}(\xi)$ is defined in \eqref{3.8}.
%%%%%%%%%%%%%%%%%%%%%%%%%%%%%%%%
\section{Results and concluding remarks}
%%%%%%%%%%%%%%%%%%%%%%%%%%%%%%%%
\begin{figure*}[hbt]
\includegraphics[width=7cm,height=5.5cm]{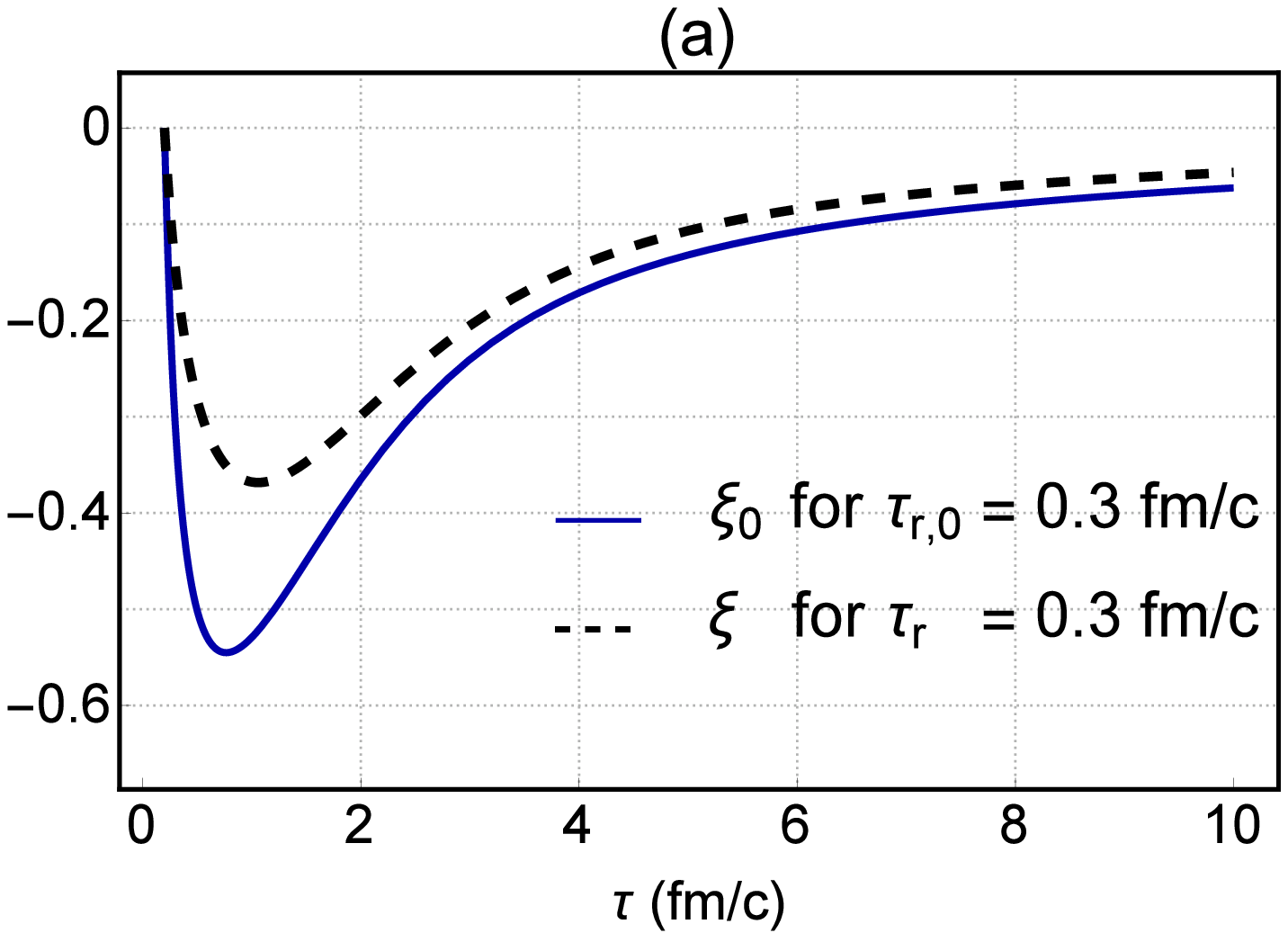}\hspace{0.8cm}
\includegraphics[width=7cm,height=5.5cm]{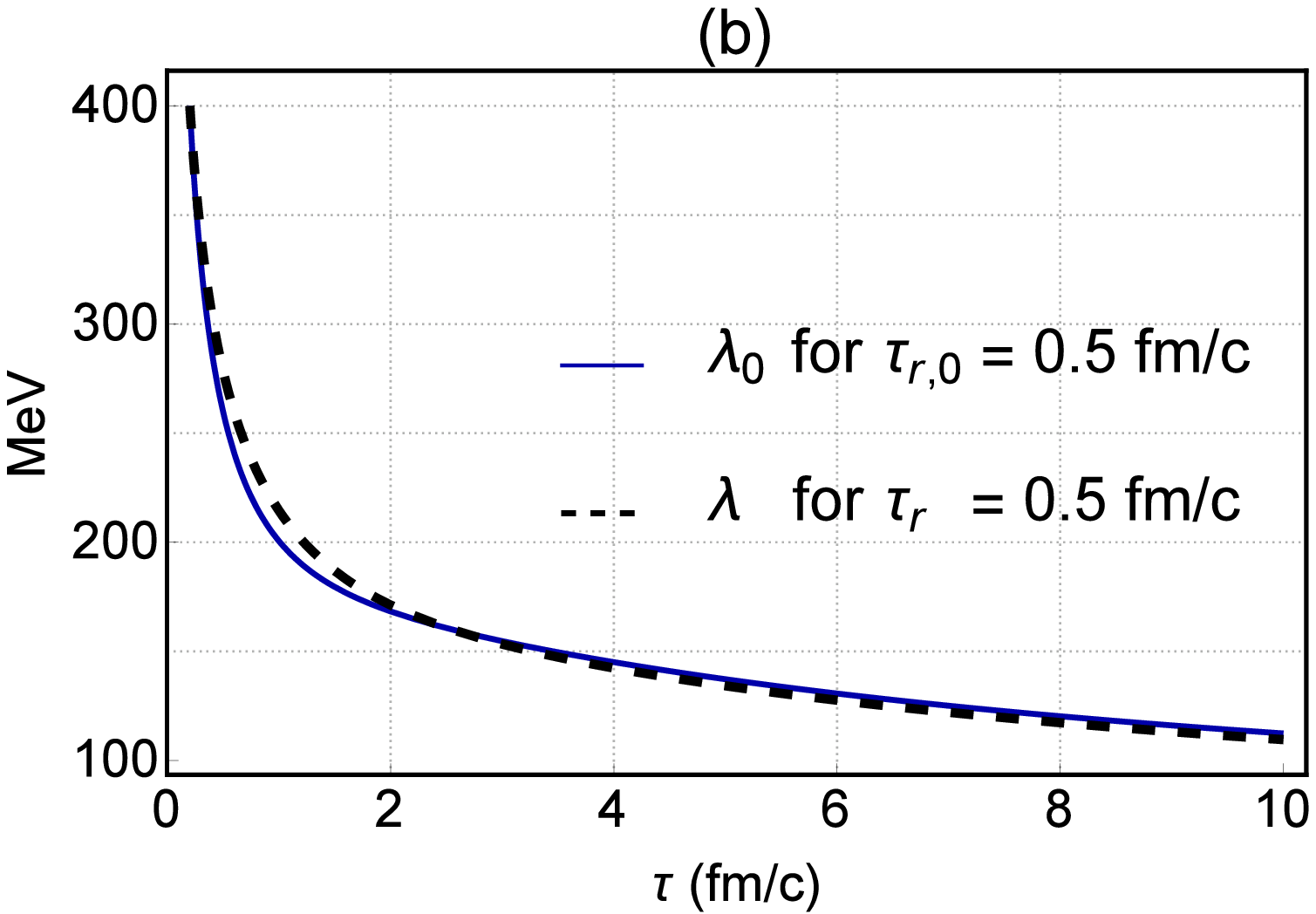}
\caption{(a) The $\tau$ dependence of the anisotropy function $\xi_{0}$ (nondissipative case) and $\xi$ (dissipative case) is plotted for relaxation times $\tau_{r,0}$ (blue solid curves) and $\tau_{r}$ (black dashed curves) equal to $0.3$ fm/c. The anisotropy parameter decreases in the early stage, and rises slowly after few fm/c. This qualitative behavior does not change once the fluid is assumed to be dissipative. (b) The $\tau$ dependence of the effective temperature $\lambda_{0}$ (nondissipative case) and $\lambda$ (dissipative case) is plotted for relaxation times $\tau_{r,0}$ (blue solid curves) and $\tau_{r}$ (black dashed curves) equal to $0.5$ fm/c. The effective temperature decreases, and a finite dissipation does not affect its qualitative behavior too much. }\label{fig-1}
\end{figure*}
\begin{figure*}[hbt]
\includegraphics[width=7cm,height=5.5cm]{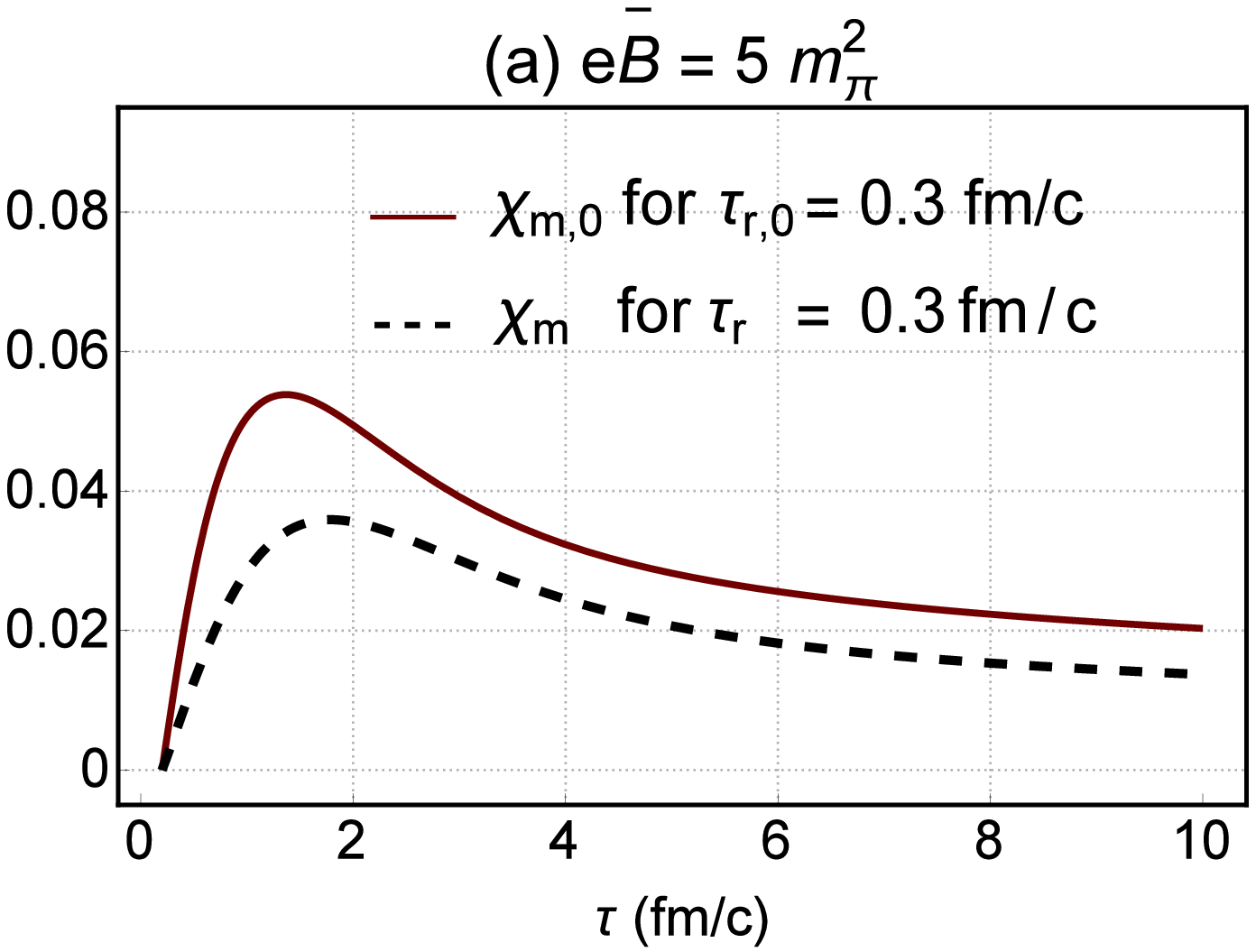}\hspace{0.8cm}
\includegraphics[width=7cm,height=5.5cm]{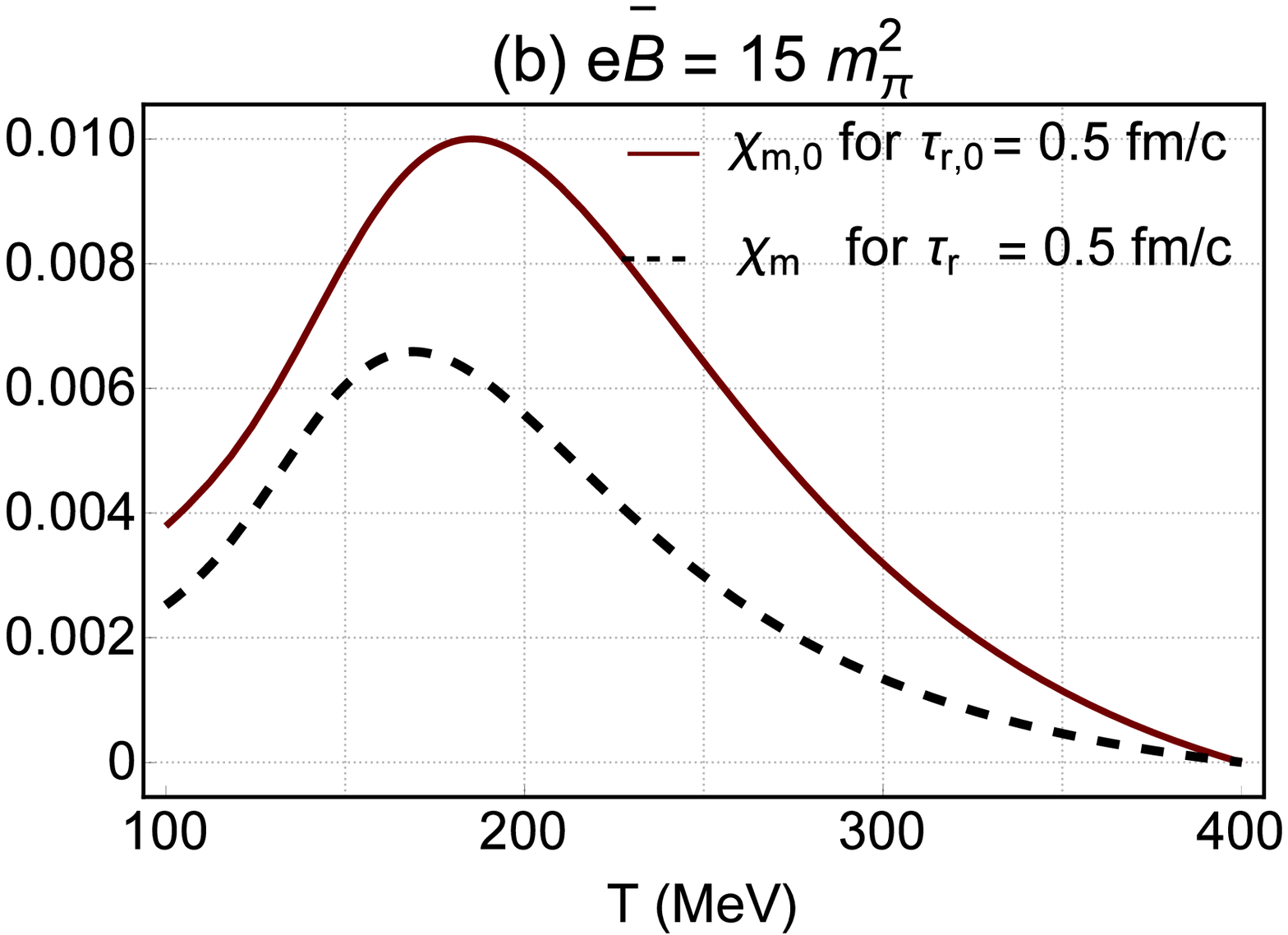}
\caption{(a) The $\tau$ dependence of the magnetic susceptibility $\chi_{m,0}$ (nondissipative case) and $\chi_{m}$ (dissipative case) is plotted for the initial magnetic field $e\bar{B}=5 m_{\pi}^2$ and relaxation times $\tau_{r,0}$ (red solid curves) and $\tau_{r}$ (black dashed curves) equal to $0.3$ fm/c. (b) Using $T=T_{0}\left(\tau_{0}/\tau\right)^{1/3}$, arising from the Bjorken solution, with the initial temperature $T_{0}=400$ MeV and the initial time $\tau_{0}=0.5$ fm/c, the $\tau$ dependence of $\chi_{m}$ is converted into its $T$ dependence. The red solid and black dashed curves demonstrate the $T$ dependence of $\chi_{m,0}$ (nondissipative case) and $\chi_{m}$ (dissipative case) for the initial magnetic field $e\bar{B}=15 m_{\pi}^2$ and relaxation times $\tau_{r,0}$ (red solid curves) and $\tau_{r}$ (black dashed curves) equal to $0.5$ fm/c, respectively.
Here, $m_{\pi}=140$ MeV. In the regime $T\in [100-200]$, the $T$ dependence of the magnetic susceptibility coincides qualitatively with lattice QCD results \cite{delia2013}. The qualitative behavior of $\chi_{m,0}$ as well as $\chi_m$ at $T\in[200,400]$ MeV can be brought in relation to the creation and the fast decay of the magnetic field at the early stages of the collision with $T\approx 400$ MeV. }\label{fig-2}
\end{figure*}
The results for the evolution of the anisotropy parameter, effective temperature, and magnetic susceptibility of the fluid in the dissipative and nondissipative cases are given in Figs. \ref{fig-1} and \ref{fig-2} for the initial values $\xi(\tau_0)= 10^{-7}$ and $\lambda(\tau_{0})=400$ MeV with $\tau_{0}=0.2$ fm/c.
The magnetic susceptibility $\chi_{m_0}$ in the nondissipative case is  determined by plugging the results for the $\tau$ dependence of $\xi_{0}$ and $\lambda_0$, determined by numerically solving (\ref{3.7}), into
\begin{eqnarray}\label{5.1}
M_0=-\frac{1}{2B}\int d\tilde{k}~\big[\left(k\cdot u\right)^{2}-3\left(k\cdot b\right)^{2}\big]f_0 =\frac{3\lambda_0^{4}}{2\pi^{2}\xi_0 B}\bigg[\left(3-\xi_0\right){\cal{R}}(\xi_0)-\frac{3}{1+\xi_0}\bigg],
\end{eqnarray}
and using $M_0=\chi_{m_0}B$. The same method is used to determine $\chi_{m}$ from $\xi$ and $\lambda$, arising from (\ref{4.15}), in the dissipative case. According to the results from Fig. \ref{fig-1}, finite dissipation strongly affects the anisotropy parameter, while the effective temperature remains almost unaffected by the inclusion of dissipation. As concerns the $\tau$ dependence of the magnetic susceptibility in Fig. \ref{fig-2}, it increases in the early stages after the collision, for $0.2<\tau<1.5$ fm/c [see Fig. \ref{fig-2}(a)], or equivalently at temperature $400>T>180$ MeV [see Fig. \ref{fig-2}(b)], and then decreases with increasing $\tau$, or equivalently with decreasing temperature. The qualitative $T$ dependence of $\chi_{m}$ in the range $T\in [100,200]$ MeV is comparable with the lattice QCD results \cite{delia2013}, where, in contrast to our result, $\chi_{m}$ continues to increase with increasing temperature. The fact that $\chi_{m}$ increases from a zero initial value at $T=400$ MeV to a maximum at $T\sim 200$ MeV is related to the creation of large magnetic fields in the early stages of the collision in this temperature regime. Let us notice that, in contrast to lattice results, the magnetic field considered in the present work is dynamical. Assuming, that in the most simplest case $B$ decays as $B\sim \tau^{-1}$, the decay of $\chi_{m}$ after reaching the maximum at $T\sim 200$ MeV becomes understandable.

\end{document}